\documentclass[conf]{new-aiaa}
\usepackage[utf8]{inputenc}
\usepackage{todonotes}

\usepackage{graphicx}
\usepackage{amsmath}
\usepackage[version=4]{mhchem}
\usepackage{siunitx}
\usepackage{subfigure}
\usepackage{longtable,tabularx}
\title{Wall modeled large-eddy simulations of flow over the Sandia Axisymmetric Transonic Bump}

\author{Rahul Agrawal \footnote{Ph.D. candidate, Dept. of Mechanical Engineering, Stanford University, and AIAA Student Member. Equal Contribution. }, Ahmed Elnahhas \footnote{Ph.D. candidate, Dept. of  Mechanical Engineering, Stanford University, and AIAA Student Member. Equal Contribution. } and Parviz Moin \footnote{Franklin P. and Caroline M. Johnson Professor, Dept. of Mechanical Engineering, Stanford University. Fellow AIAA. } }
\affil{Center for Turbulence Research, Stanford University, California, United States of America - 94305}

\begin{document}

\maketitle

\begin{abstract}

Wall-modeled large-eddy simulations (WMLES) are conducted for the flow over the Sandia Axisymmetric Transonic Bump (ATB) at bump chord Reynolds number, $Re_c = 1 \times 10^6 $ and Mach number, $Ma_{ref} = 0.875$. Utilizing various subgrid-scale and wall models, comparisons are made between the simulations and experiments for quantities of engineering interest, such as the skin friction coefficient ($C_f$) and wall pressure coefficient ($C_p$) data. Favorable agreements between the simulations and the experiments are achieved, with the simulations predicting the shock location, size, and strength of the separation bubble with reasonable accuracy. In particular, it was established that the realism of the incoming turbulent boundary layer onto the bump is very important for the correct and robust prediction of the skin friction in the upstream region of the bump. Furthermore, a resolution limitation on the predictive capability of the equilibrium wall model in the favorable pressure gradient region is established. Finally, preliminary results utilizing a pressure-gradient inclusive modification to the equilibrium wall model predict the skin friction within the separation bubble at affordable resolutions.

\end{abstract}

\section{Nomenclature}

{\renewcommand\arraystretch{1.0}
\noindent\begin{longtable*}{@{}l @{\quad=\quad} l@{}}
$C_p$ & pressure coefficient \\
$C_f$ & skin-friction coefficient \\
$p_{ref}$ & reference pressure in simulation domain \\
$\gamma$ & heat capacity ratio \\
$\tau_w$ & wall stress \\
$T$ & temperature \\
$E$ & sum of internal and kinetic energies \\
$u_i$ & velocity components \\
$N_{cv}$ & number of control volumes in simulation domain\\
$S^{d}_{ij}$ & deviatoric component of the strain-rate tensor \\
$L_{ij}$ &  resolved turbulent stresses\\
$C_{s}$ &  model coefficient of dynamic Smagorinsky model\\
$C_{ij}$ &  model coefficients of dynamic tensor-coefficient Smagorinsky model\\
$\nu$ & kinematic viscosity \\
$\rho$ & density \\
$y^+$ & viscous length scale \\
$U_{e}$ & freestream velocity \\
$Re_c$ & Reynolds number based on bump length \\
$Ma_{ref}$ & Mach number \\
$\delta_{exp}$ & boundary layer thickness based on experimental velocity profiles\\
\end{longtable*}}

\section{Introduction}

The flow over a commercial aircraft is characterized by the presence of complex flow phenomena ranging from three-dimensional juncture flows \cite{lee2019overflow,lozano2021performance}, smooth-body separation \cite{rumsey2019overview}, and transonic flow phenomena including shock-induced boundary layer separation \cite{vassberg2008development}. Accurately predicting all these flow phenomena using a unified framework is necessary to determine the lift and drag forces within the required industry tolerances. While Reynolds-averaged Navier-Stokes (RANS) simulations have been the primary simulation paradigm, their stagnation in predictive accuracy has led to the emergence of scale-resolving simulations, such as large-eddy simulations (LES), a more accurate alternative. However, true LES near the wall is prohibitively expensive for engineering flows of interest due to the scaling of the size of the energy-carrying eddies as the wall is approached \cite{choi2012grid}. This prohibitive cost scaling can be overcome by replacing the near-wall region with a reduced-order model, i.e., a wall model, which is the basis for the wall-modeled large eddy simulation paradigm (WMLES) \cite{bose2018wall}. WMLES has been successfully used to accurately predict two of the three complex flow phenomena highlighted above. In particular, are the accurate predictions of the three-dimensional juncture flows using the NASA juncture flow experiment \cite{lozano2020prediction,lozano2022performance}, and smooth-body separation using the Boeing bump experiment \cite{agrawal2022non}. In this study, we examine the predictive capabilities of state-of-the-art WMLES technology in predicting transonic shock-induced separation using the Sandia ATB experiment \cite{lynch2019revisiting,lynch2020cfd,beresh2020cfd,lynch2023experimental}. 

Earlier, Bachalo and Johnson \cite{bachalo1986transonic} conducted an experimental investigation of a canonical flow that isolates the transonic shock-induced flow separation in conjunction with smooth-body separation flow physics. In this experiment, an axisymmetric model with a spherical bump is placed in a flow with a transonic Mach number. The turbulent boundary layer that develops over the model's surface upstream of the bump interacts with the local supersonic flow over the bump, leading to the detachment of the boundary layer and flow separation. The axisymmetric flow model provides a flow field free from side wall interference and three-dimensional effects. The results of this experiment were used to calibrate industry-standard low-fidelity RANS models (e.g., SST model \cite{menter1994two}). However, these models systematically mispredict critical aspects of the flow, such as the separation bubble size and the turbulent shear stress \cite{debonis2015test}. Subsequently, scale-resolving wall-modeled large eddy simulations \cite{spalart2017large} and wall-resolved simulations \cite{uzun2019wall} calculations have been performed, with more accurate results but persistent systematic mispredictions in the mean velocity profiles and turbulent stresses remain. The lack of reference experimental data such as inflow freestream turbulence intensity, second-order turbulence statistics, unsteady shock location measurements, and skin-friction data, accompanied by uncertainty in model geometry, unquantified wind-tunnel effects, and the prohibitively large Reynolds number of the experiment for manageable high-fidelity numerical simulations, led to a new experimental campaign at Sandia National Labs that aimed to address these issues \cite{lynch2019revisiting,lynch2020cfd,beresh2020cfd,lynch2023experimental}.

The experiment was conducted at the same transonic Mach number but at a lower Reynolds number to decrease the cost of numerical studies. A blind CFD challenge was conducted between January 2020 and January 2021, where participants submitted simulation results using different computational approaches such as RANS and scale-resolving LES \cite{riley2021rans, rahmani2022large, gupta2021shock}. The main evaluation metrics of the success of the models, based on engineering quantities of interest (QoIs), included the prediction of the mean velocity and Reynolds stress profiles, mean separation and reattachment locations, mean surface pressure measurements, as well as wall-shear stress distribution and unsteady turbulent-induced pressure loading.

A key QoI is the skin-friction coefficient distribution along the model axial extent, as it illustrates the ability of the modeling paradigm to predict the stress at the wall, along with the separation and reattachment points, correctly. The previous attempts at simulating this flow led to the following observations. RANS entries could reasonably predict the shock location, but the reattachment location was incorrect \cite{riley2021rans}. Furthermore, the skin friction coefficient before the shock depended on the choice of the RANS model. A scale-resolving, WMLES calculation predicted the reattached skin friction correctly but overestimated the upstream skin friction and mispredicted the shock location \cite{riley2021rans}. Finally, both Refs. \cite{rahmani2022large, gupta2021shock} mispredicted the skin-friction coefficient in the majority of the flow. 

In this work, we utilize the state-of-the-art WMLES in the charLES framework \cite{bres2018large} which relies on non-dissipative numerics, dynamic subgrid-scale models, equilibrium wall models, and unstructured meshing to predict the flow around the Sandia ATB. The findings in this work are a follow-up to those reported in Ref. \cite{elnahhaswall}. The results are compared to the recently released set of experimental measurements \cite{lynch2023experimental}, which differ from the dataset used for comparisons in the aforementioned computational studies, due to experimental reporting errors upstream of the bump. The rest of the paper is organized as follows. Section III discusses the choice of subgrid-scale and wall models used in the majority of this work. Some details of the numerical solver are provided in Section IV. Section V presents the experimental geometry as well as the mesh size and resolution. Section VI describes the computational boundary and inlet conditions. The results with the dynamic subgrid-scale models and the equilibrium wall model are presented in Section VII. Section VIII addresses the effect of the upstream history of the incoming boundary layer onto the bump. Section IX provides a brief description of a recently developed pressure-gradient inclusive wall model along with preliminary results of its application on the Sandia ATB. Finally, conclusions are presented in Section X. 

\section{Governing equations and modeling approaches }
In LES, the coarse-grained, large-scale quantities are defined by filtering the velocity and pressure fields. Denoting the grid-filter kernel operator by $\mathcal{G}$, the large scale quantity, $\tilde{f}$, can be evaluated from the total field, $f$ as  
\begin{equation}
\tilde{f}(x)    = \int \mathcal{G}(x,x') f(x') dx'
\end{equation}
where the integral is extended \textcolor{black}{over} the entire simulation domain. Similarly, we define $\overline{f}$ to be a Favre averaged variable as,
\begin{equation}
    \overline{f} = \frac{ \widetilde{f\rho}}{\tilde{\rho }}
\end{equation}
More details on LES formalism can be found in previous studies \cite{germano1991dynamic,ghosal1995dynamic,moin1991dynamic}.  

With these definitions, for a compressible turbulent flow (of internal energy $e$, density $\rho$,  temperature $T$, viscosity $\mu(T)$ and thermal conductivity $\kappa(T)$ ) and velocity vector $\vec{u} \; = \; \{ u_1, \;u_2, \;u_3\}$, the  governing equations can be obtained by applying the aforementioned filters to the Navier-Stokes equations to arrive at 

\begin{equation}
\frac{\partial \tilde{\rho} }{\partial t }
+ \frac{\partial (\tilde{\rho} \; \overline{u}_i )}{\partial x_i} = 0  
\end{equation}

\begin{equation}
\frac{\partial  (\tilde{\rho} \; \overline{u}_i )}{\partial t }+\frac{\partial (\tilde{\rho} \; \overline{u}_j \;  \overline{u}_i )}{\partial x_j } =- \frac{\partial \tilde{p}}{\partial x_i } + \frac{\partial (\mu \overline{S^d}_{ij} ) }{\partial x_j } -\frac{\partial \tau^{sgs}_{ij}}{\partial x_j} ,
\end{equation}

and

\begin{equation}
\frac{\partial  \tilde{E}}{\partial t }+\frac{\partial (\tilde{E} \; \overline{u}_j)   }{\partial x_j } =- \frac{\partial (\tilde{p}\; \overline{u_i}) }{\partial x_i } + \frac{\partial   (\mu \overline{S^d_{ij}} \overline{u_i} )}{\partial x_j } -\frac{\partial (\tau^{sgs}_{ij} \overline{u_i})  }{\partial x_j} - \frac{\partial Q_j^{sgs} }{\partial x_j } + \frac{ \partial }{\partial x_j } ( \kappa \frac{\partial \tilde{T}}{\partial x_j }) ,
\end{equation}
where $\tilde{E} =\tilde{\rho} \; \overline{e} + 0.5 \;  \tilde{\rho}\;  \overline{u_i}\;  \overline{u_i}$ is the sum of the resolved internal and kinetic energies and $\overline{S^d}_{ij}$ is the deviatoric part of the resolved strain-rate tensor. A power law (with the exponent 0.75) is used to determine the temperature dependence of molecular viscosity, and the corresponding thermal conductivity is evaluated using a constant molecular Prandtl number approximation ($Pr = 0.7$).

Note that  $\tau^{sgs}_{ij} = \tilde{\rho } (\overline{u_i u_j} - \overline{u}_j \;  \overline{u}_i) $  and  $Q_j^{sgs} = \tilde{\rho} (\overline{e u_j} - \overline{e}\; \overline{u_j} ) $ are the subgrid stress and heat flux respectively which require modeling closure. The isotropic component of the subgrid stress is often absorbed into pressure, which leads to a pseudo-pressure field. In the following subsection, we discuss the two subgrid-scale models \cite{germano1991dynamic, agrawal2022non} used for comparative study in this work. The subgrid heat flux is modeled using the constant turbulent Prandtl number approximation ($Pr_t = 0.9$) based on the dissipative component of the subgrid-stress tensor. 

\subsection{Choice of the subgrid-scale model }

\subsubsection{Dynamic Smagorinsky Model (DSM)}
Germano et al. \cite{germano1991dynamic} introduced the idea of test filtering in LES, to relate the resolved turbulent stresses to the modeled subgrid-stresses through a dynamical closure procedure. The subgrid-stress tensor is modeled as, 
\begin{equation}
    \tau_{ij} = -2 (C_s \Delta^2 ) |S|  S_{ij},
    \label{eqn:eddyvismodels}
\end{equation}
with the model coefficient being determined dynamically as,  
\begin{equation}
     (C_s \Delta)^2 =  \frac{\langle L_{ij} M_{ij}\rangle}{\langle 2 M_{ij} M_{ij}\rangle}.
    \label{eqn:dsm}
\end{equation}
where $\langle \cdot \rangle$ is the spatio-temporal averaging operator, $L_{ij} = - \widehat{\overline{u}_i\overline{u}_j} + \widehat{\overline{u}_i} \; \widehat{\overline{u}_j}$ ($\widehat{(\cdot)}$ denotes test-filter operation) is the resolved stress tensor. Similarly, the $M_{ij}$ tensor is a measure of the modeled stresses between the grid and the test filter levels, and is given as 
\begin{equation}
    M_{ij} =  \left( {\frac{\widehat{\Delta}^2}{\Delta^2}} \widehat{|S|}\widehat{S_{ij}} - \widehat{|S|S_{ij}}\right), 
    \label{eqn:mij}
\end{equation}

\subsubsection{Dynamic Tensor-coefficient Smagorinsky model (DTCSM)}
For LES of flows with mean flow-driven anisotropy (for example, turbulent boundary layers experiencing mean pressure gradients), the assumption of a scalar model coefficient that is agnostic to the direction of anisotropy may be challenged. 
Agrawal et al. \cite{agrawal2022non} developed a non-Boussinesq subgrid-scale model (hereby abbreviated as DTCSM) in which the model coefficients are allowed to vary tensorially. Further, the realizability constraints as posed by the Navier-Stokes equations, and the dynamic procedure \cite{germano1991dynamic} are invoked to model all of the model coefficients dynamically. Mathematically, this model can be written as 
\begin{equation}
    \tau_{ij}^{sgs} - \frac{\tau^{sgs}_{kk}  }{3} \delta_{ij} = - (C_{ik}\overline{S}_{kj} + C_{jk}\overline{S}_{ki} )|\overline{S}| \tilde{\rho} \Delta^2. 
\label{eqn:dtcsm0}
\end{equation}
Agrawal et al. \cite{agrawal2022non} argued that for this model to satisfy its tracelessness property, the coefficients can be related to each other as 
\begin{equation}
    C_{11} = C_{22} = C_{33} ~; \qquad
    C_{ij} = - C_{ji} \quad (j \neq i) ~.    
    \label{eqn:dtcsm1}
\end{equation}
These realizability constraints reduce the number of independent coefficients from nine to four. Applying the Germano identity to this model form, the following equations are derived and solved in the least squares sense to dynamically compute all model coefficients (the reader is referred to Ref. \cite{agrawal2022non} for more details).

\subsection{Choice of the wall model }
In the majority of this work, an algebraic formulation of the equilibrium wall-stress model (hereby abbreviated as EQWM), in which the assumed mean velocity profile is $C^1$ continuous between the viscous sublayer and the logarithmic layer, is used. Details of the compressible formulation of the equilibrium wall model can be found in \cite{lehmkuhl2018large}. Unlike some of the previously reported studies that use higher off-wall matching points to avoid numerical inaccuracies \cite{kawai2012wall}, first-point matching has been used in the present study. With this solver (that utilizes Voronoi, hexagonally closed-packed cells), on performing simulations of a turbulent channel flow at $Re_{\tau} = 4200$ with a typical WMLES resolution of 20 points per channel half-height, no log-layer mismatch has been observed.

\section{Numerical Solver details}
The simulations presented in this work were performed using an explicit, unstructured, finite-volume solver for the compressible Navier-Stokes equations (charLES). This code is 2\textsuperscript{nd}-order accurate in space, and 3\textsuperscript{rd}-order accurate in time, and utilizes Voronoi-based grids. More details of the solver and validation cases can be found in Refs. \cite{bres2018large,goc2021large}. 

\section{Description of computational domain, and meshing topology} 
\label{sec:mesh}
\begin{figure}[!ht]
    \centering
    \includegraphics[width=1\textwidth]{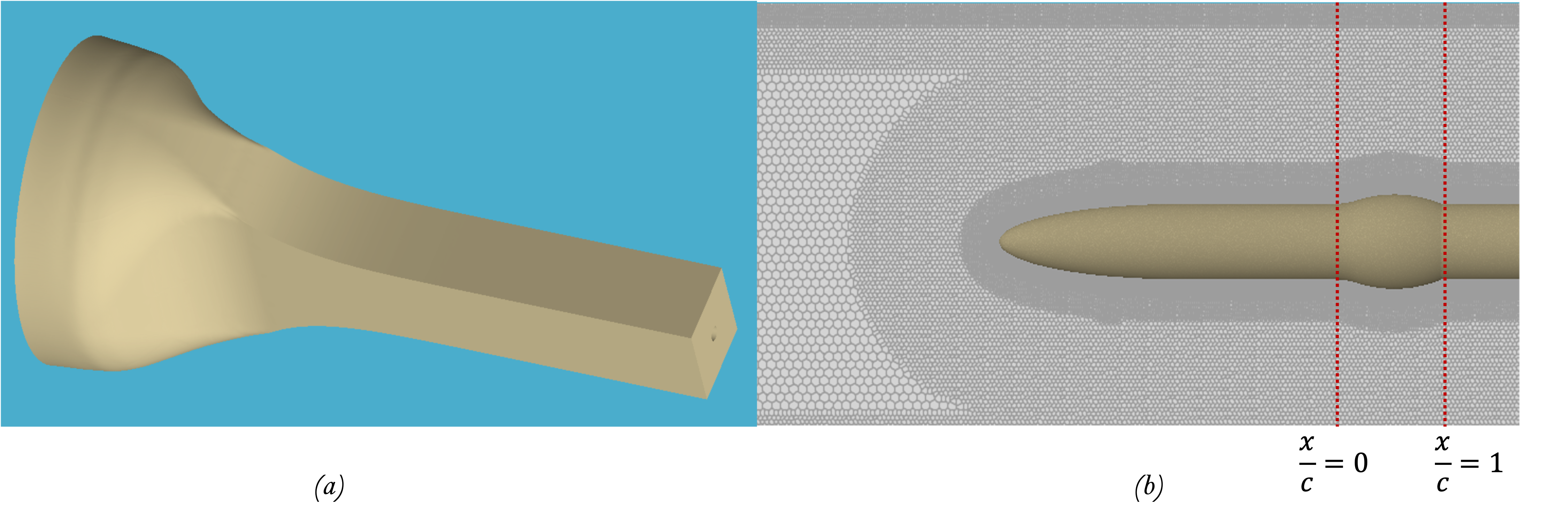}
    \caption{ (a) A representation of the wind tunnel wall geometry as simulated in the present work and (b) a view of the grid refinement procedure near the bump wall. The denser colors represent the near-wall grid refinement.  Note that the sub-figure in the right panel is representative of the coarse mesh studied in this work.  }
    \label{fig:Geometry/Grid}
\end{figure}

The simulated domain encompasses an accurate representation of the axisymmetric model (including the nose, and the small step upstream of the bump) and the full wind tunnel geometry (available in Lynch et al. \cite{lynch2020cfd} and the references therein). Figure \ref{fig:Geometry/Grid}(a) illustrates the wind tunnel domain. Figure \ref{fig:Geometry/Grid}(b) showcases the grid refinement strategy, where the grid is linearly, and homothetically refined  near the wall of the axisymmetric model. Physically, this linear refinement is consistent with the scaling of the energy-containing eddies near the wall in wall-bounded turbulence. Unlike some of the existing studies \cite{iyer2017wall,riley2021rans,gupta2021shock} of this flow, a unique feature of these meshing practices is that they are completely agnostic to the presence of shocks in the flow. No special treatments/refinements along the streamwise direction were pursued around the shock and separation locations. Further, to draw more concrete conclusions about the predictive capabilities of wall-modeled LES across different Mach number and Reynolds number regimes, the outer scaling of these grid resolutions ($\delta/\Delta$ where $\delta$ is the boundary layer thickness, and $\Delta$ is the cell size) was approximately matched to those used in Refs. \cite{agrawal2022non,agrawalarb2022,whitmorebump} which focused on a subsonic smooth body separation problem of the Boeing speed bump.  

    The grid resolutions employed to converge the present simulations are quantified using the number of grid points within the experimental boundary layer thickness upstream of the favorable pressure gradient region of the bump. These values and the total number of grid points for each simulation, $N_{cv}$, are reported in Table \ref{table:reshump}. It is important to note that while these values are larger than ones reported in other studies, the present work simulates the full azimuthal extent of the axisymmetric geometry, unlike most previous RANS and WMLES studies. For example, Ref. \cite{riley2021rans} simulated a quarter of the experimental domain ($90^o$ sector) with resolutions comparable to the present work. Similarly, Ref. \cite{rahmani2022large} simulated only a quarter of the domain, and did not include the entire wind tunnel, which requires approximately 40 million degrees of freedom. Finally, while Ref. \cite{gupta2021shock} used O(200) MCVs for a full axisymmetric geometry, they removed the upstream portion (for example, the nose and the backward-facing step on the bump wall) of the geometry such that their inlet was positioned at $x/c = -0.78$ where the flow develops into a zero  pressure gradient boundary layer. This greatly reduced the number of grid points required, but  the results mispredicted $C_f \; \mathrm{and} \; C_p $  compared to the experiments within the separation and recovery regions of the flow.

\begin{table}
\caption{Mesh parameters for wall-modeled LES of the Sandia transonic bump at $Re_c = 1\times 10^6 \; \mathrm{and} \; Ma = 0.875$.}
\label{table:reshump}

\begin{center}
    \begin{tabular}{ p{3.5cm}p{3.5cm}p{4.5cm}}
Mesh & $N_{cv}$ & $\delta_{Exp} / \Delta$ (at $x/c = -0.78$)  \\
\hline\noalign{\vspace{3pt}}
Coarse  & $60$ Mil. & $9$ \\
Medium  & $200$ Mil. & $12$  \\
Fine    & $780$ Mil. & $16$ \\
\end{tabular}
\end{center}

\end{table}

\section{Computational boundary and inlet conditions }

\subsection{Boundary Conditions}
The inlet boundary condition for the current simulations is set to be a plug flow at $x/c = -34 $ where $c$ is the length of the bump. The outlet (at $x/c = 6 $) is treated using the non-reflecting characteristic boundary condition with constant pressure \cite{poinsot1992boundary}. The wind tunnel walls are treated as inviscid; in a separate sequence of calculations (not shown), the authors have verified that the wind tunnel boundary conditions have a relatively weak effect on the quantities of interest. The algebraic form of the equilibrium wall model is applied to the axisymmetric model containing the bump.

\subsection{Inlet conditions }
In the majority of this work, we simulate the experimental geometry along with the wind tunnel to faithfully reproduce the experiment \cite{lynch2020cfd}. Further, the experiments also provide a combination of necessary flow details such as the stagnation values of the temperature, pressure, and Mach number measurements at two stations, $x/c = -6.31$ and $x/c = 0$. Assuming isentropic conditions before the flow experiences a shock (approximately at $x/c \sim 0.65$ as per Ref. \cite{lynch2023experimental}), and using the mass, momentum, and energy conservation equations, inlet conditions can be derived for the wind tunnel. It has also been verified that the pressure at the wind tunnel wall at $x/c = 0$ is within one percent of the experimentally recorded pressure at the same location, thus reasonably complying with the requirements of the CFD Validation challenge held in AIAA SciTech Forum 2021.

\section{Results }
In this section, we discuss the WMLES results for the wall-modeled large-eddy simulations of the Sandia ATB using both the two aforementioned subgrid-scale models and the equilibrium wall model. The quantities of interest for this flow are the skin friction and pressure coefficients over the streamwise extent of the bump. These quantities are defined as, 
\begin{equation}
    C_f = \frac{\tau_{w}}{ 1/2 \gamma  p_{ref} Ma_{ref}^2 }  \hspace{5pt} \mathrm{and} \hspace{5pt}  C_p = \frac{ p - p_{\rm ref}}{ 1/2 \gamma  p_{ref} Ma_{ref}^2 } .
\end{equation}
where $\gamma$, $\tau_{w}$, $p$, and $p_{\rm ref}$ are the heat capacity ratio, mean wall stress, wall pressure, and reference pressure, respectively. The reference pressure and Mach number are defined as the bump wall pressure and Mach number at the beginning of the bump, $x/c = 0$ (refer to Ref. \cite{lynch2020cfd} for more details). 

The results are organized as follows. A grid refinement sweep is completed for the standard WMLES modeling practices (DSM and EQWM) to establish grid sensitivities, and provide a baseline expectation of the predictive capabilities of WMLES. Next, the subgrid-scale model sensitivities are investigated by using DTCSM instead of DSM. The equilibrium wall model is used in this section, based on the prior observations of Park et al. \cite{park2017wall} in which the non-equilibrium wall stress model did not improve the predictions of the skin friction in the flow over NASA wall-mounted bump.

\begin{figure}[!ht]
    \centering
    \includegraphics[width=0.7\textwidth]{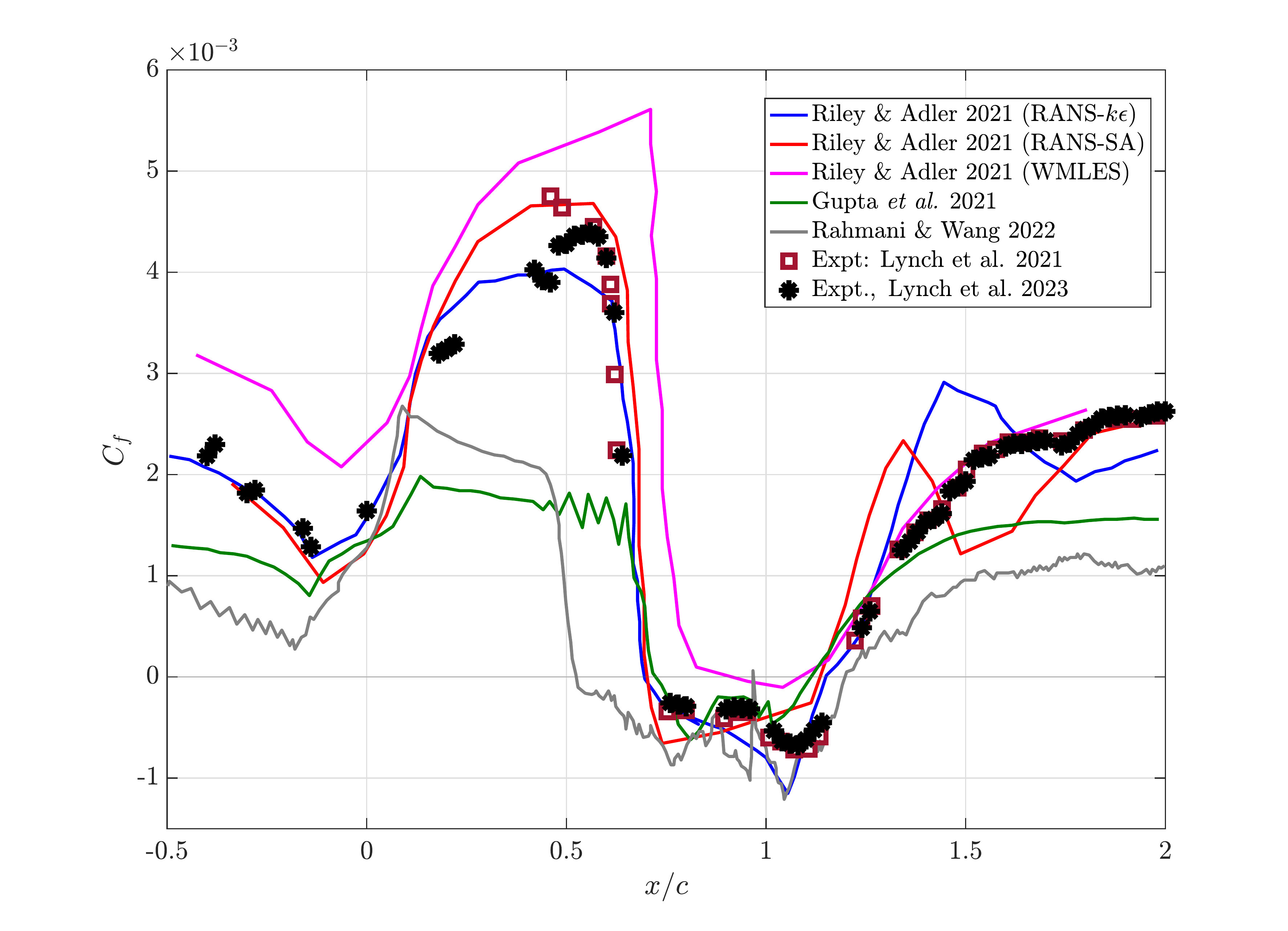}
    \caption{  The distribution of skin friction coefficient for the existing CFD results for the flow over the Sandia ATB at $Re_c = 1 \times 10^6 \, \mathrm{and} \;  Ma = 0.875$. The existing CFD simulations were obtained from Refs. \cite{riley2021rans,gupta2021shock,rahmani2022large}. The originally reported experimental measurements, and the revised measurements have both been plotted.}
    
    \label{fig:OldCfResults}
\end{figure}

Figure \ref{fig:OldCfResults} has been adapted from various existing studies \cite{gupta2021shock,rahmani2022large,riley2021rans} of this flow configuration; comparisons between the predicted $C_f$ and the original and since revised experimental data have been made. It is noted that the revised experimental results were not available at the time when the simulations were reported. However, it is fairly apparent that the previous attempts inaccurately predict the skin friction in one flow region or the other. For instance, the RANS simulations of Riley and Adler \cite{riley2021rans} are fairly accurate until the region of flow recovery. Their wall-modeled LES overpredicts the $C_f$, in the upstream, and favorable pressure gradient regions, with a diminished separation bubble size. Finally, the results from Gupta et al. \cite{gupta2021shock} and Rahmani and Wang \cite{rahmani2022large} show unphysical oscillations in skin friction.

\subsection{WMLES using DSM and EQWM on experimental geometry}
\begin{figure}[!ht]
    \centering
    \subfigure[]{
    \includegraphics[width=0.48\textwidth]{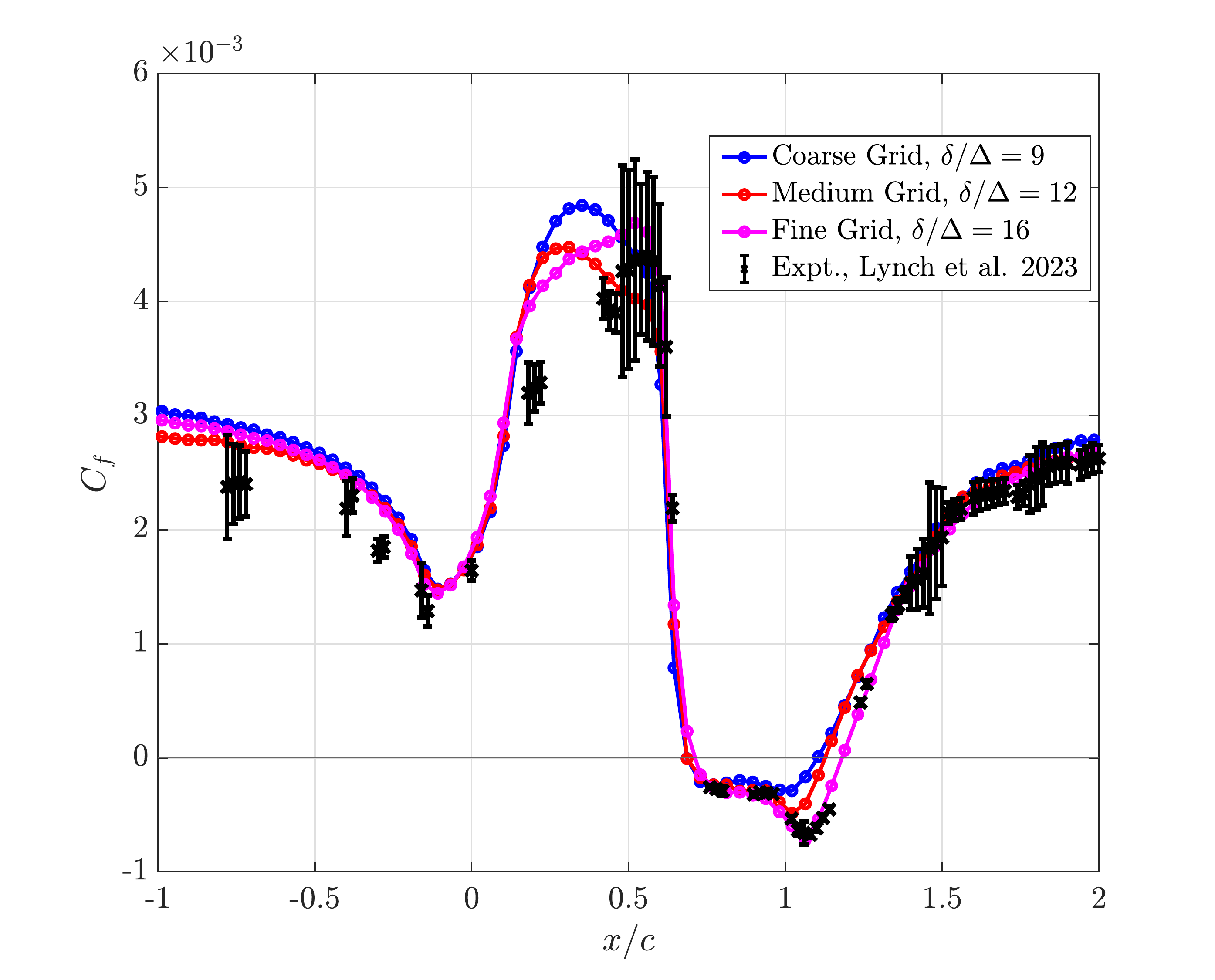}}
    \subfigure[]{
    \includegraphics[width=0.48\textwidth]{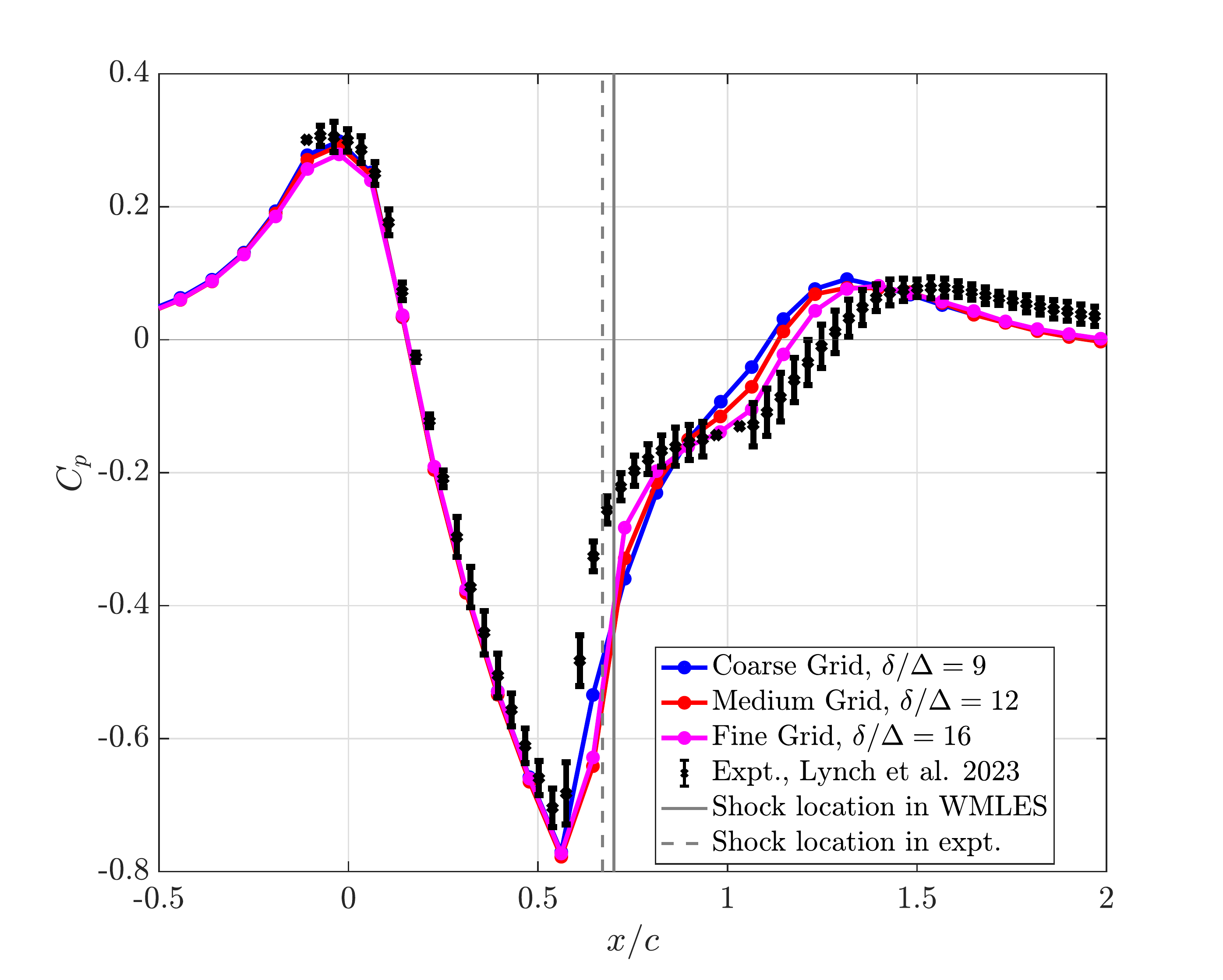}} 
    \caption{  The distribution of (a) Skin friction and (b) wall pressure coefficients for the wall modeled LES using the dynamic Smagorinsky subgrid-scale model (DSM) and equilibrium wall model of the transonic Sandia ATB at $Re_c = 1 \times 10^6 \, \mathrm{and} \;  Ma = 0.875$. Note that the error bars for the skin friction refer to the uncertainty in the experimental measurements.    }
    
    \label{fig:CfCpDSMResults}
\end{figure}

Here, a grid convergence study is reported using the dynamic Smagorinsky subgrid-scale model (DSM), and the equilibrium wall model. In Fig. \ref{fig:CfCpDSMResults}, the skin friction and wall pressure coefficient distributions along the streamwise direction are compared with the experimental data obtained in 2021 \cite{lynch2020cfd} and 2023 \cite{lynch2023experimental}. A few observations can be made; the upstream (of the bump) prediction of the skin friction is almost grid-insensitive and does not agree with the experiment, thus indicating errors in the development of the turbulent boundary layer upstream of the bump. Once the flow reaches the bump ($x/c \sim 0$), it experiences a favorable pressure gradient region marked by the increase in skin friction. Consistently, on all grids, the skin friction predicted by the simulation is higher than the measurements in the experiment (at $x/c \sim 0.2$). Since the flow experiences a favorable pressure gradient region, the Reynolds shear stresses decrease, with a possibility of flow relaminarization. 

As the flow nears the apex of the bump ($x/c \sim 0.5$), an apparent non-monotonicity is visible in the WMLES predictions. The finest grid behaves differently from the coarser counterparts in that the $C_f$ increases between $0.2 < x/c < 0.6$. It is hypothesized that as the favorable pressure gradient reduces the Reynolds shear stresses, and thins the local boundary layer, a more refined calculation is needed to capture the near-wall behavior of the flow; indeed, the finest grid qualitatively matches the experimental trend in the increase in $C_f$ between $0.2 < x/c < 0.6$. The wall pressure distribution suggests a monotonic convergence towards the experiment in all regions of the flow, however, the shock location seems to be slightly delayed (downstream of the true location) and the separated region is slightly smaller compared to the experiment. 

\begin{figure}[!ht]
    \centering
\includegraphics[width=0.6\textwidth]{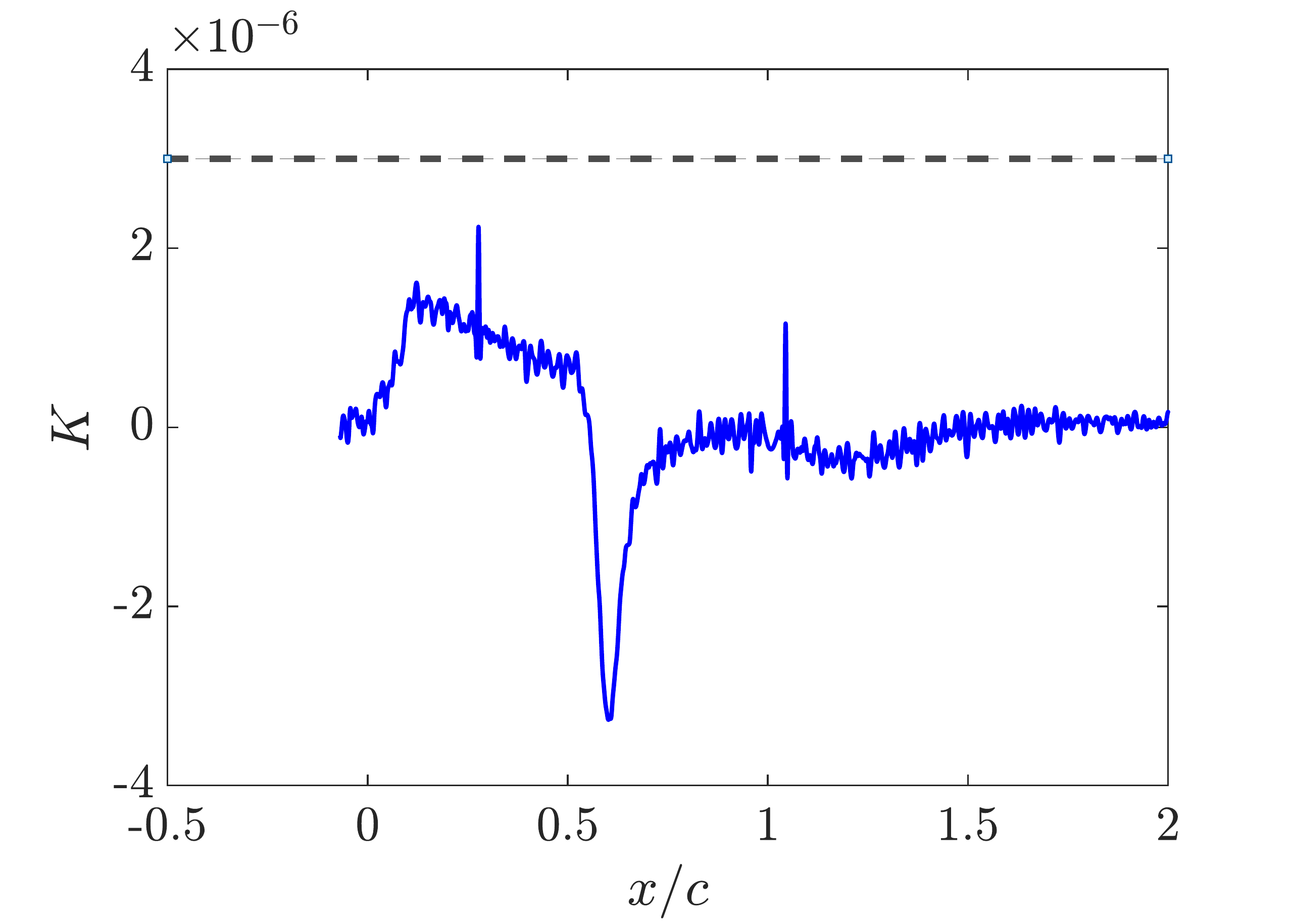}
    \caption{  The distribution of the Launder parameter,  $\kappa = \frac{\nu}{U^2_e} \frac{\partial U_e}{\partial x}$ 
 in the streamwise direction for the flow over the Sandia ATB. The dotted line is the commonly used \cite{uzun2017wall} threshold for determining the effects related to flow relaminarization.  }
    \label{fig:Relam}
\end{figure}

The drop in the Reynolds stresses and/or the partial relaminarization of the flow in the favorable pressure gradient region can not be accurately captured with a fully turbulent wall model (such as the EQWM) on relatively coarse grids. Figure \ref{fig:Relam} shows the streamwise development of the Launder parameter, $\kappa = \frac{\nu}{U^2_e} \frac{\partial U_e}{\partial x}$ \cite{jones1972prediction} obtained from the experimental pressure data. While the experimental data remains below the commonly quoted threshold value for relaminarization of $3 \times 10^{-6} $, it is large enough to experience non-trivial modifications to the mean flow caused by a large drop in the Reynolds stresses and/or partial relaminarization \cite{finnicum1988effect}. This is corroborated with the newest experimental measurements \cite{lynch2023experimental} which show an $80\%$ decrease in the Reynolds normal stress in this region of the flow. 

\subsection{WMLES with DTCSM and EQWM on experimental geometry}
\begin{figure}[!ht]
    \centering
    \subfigure[]{
    \includegraphics[width=0.47\textwidth]{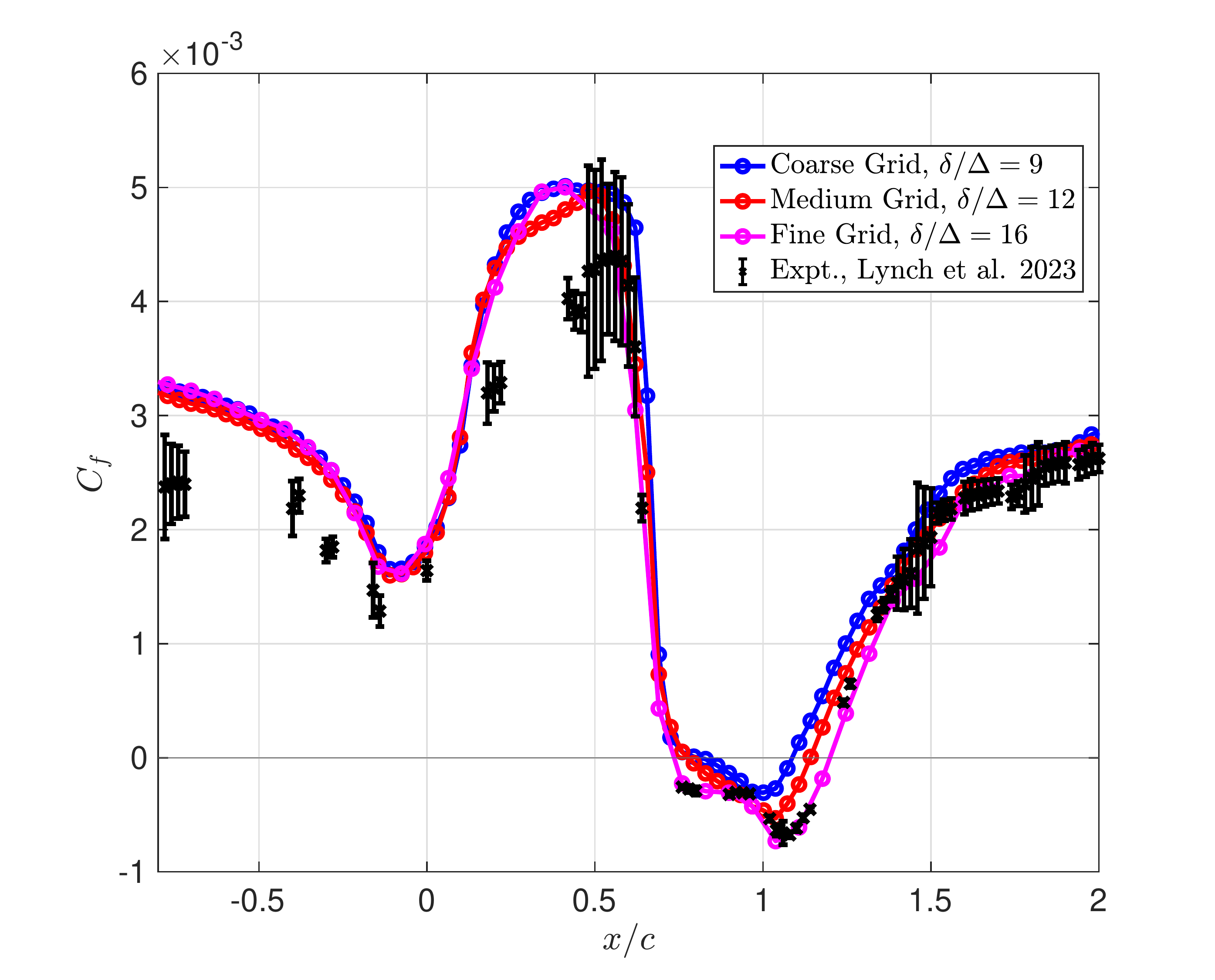}}
    \subfigure[]{
    \includegraphics[width=0.49\textwidth]{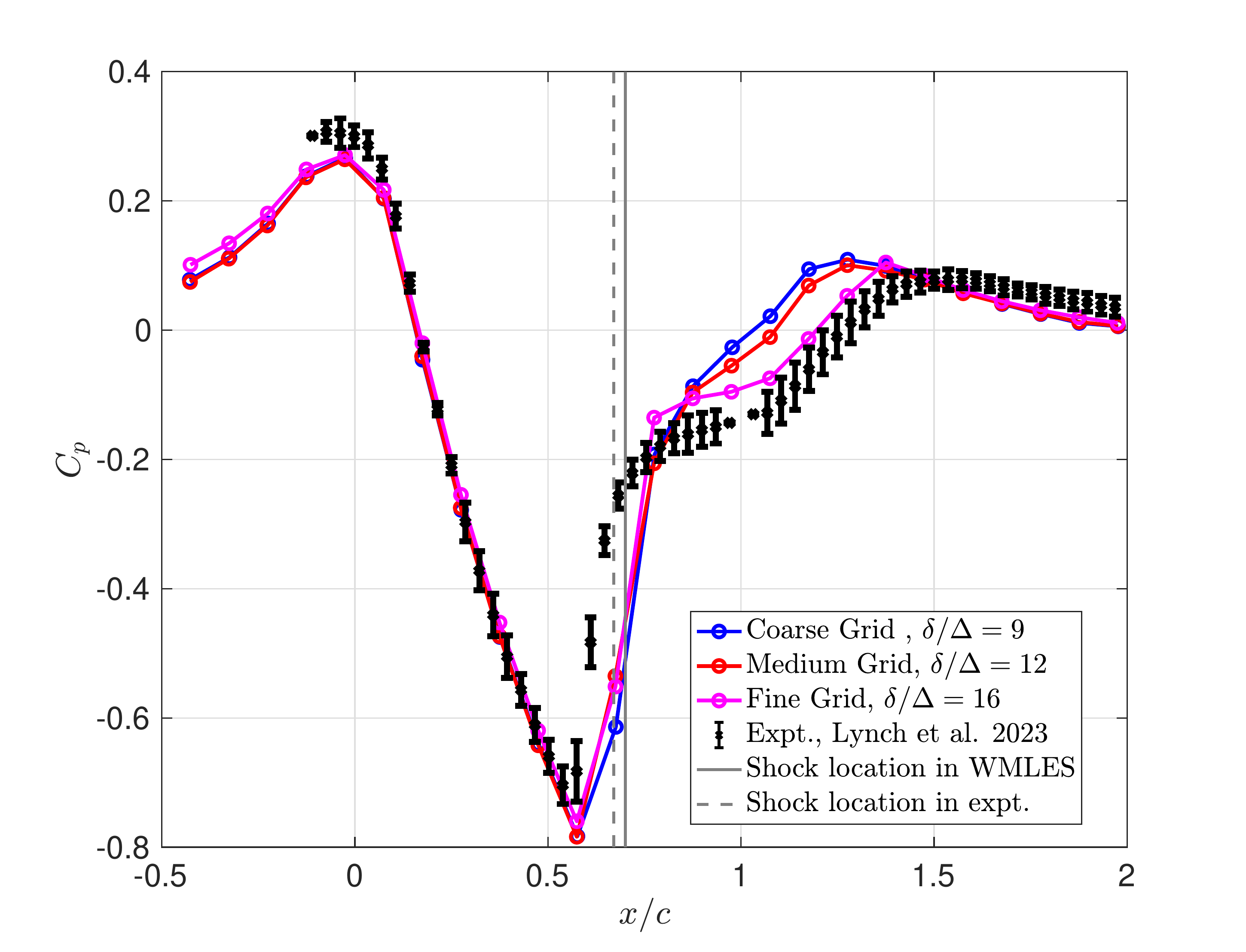}} 
    \caption{  The distribution of (a) Skin friction and (b) wall pressure coefficients for the wall modeled LES using the dynamic tensor-coefficient Smagorinsky subgrid-scale model (DTCSM) and the equilibrium wall model of the Sandia ATB at $Re_c = 1 \times 10^6 \, \mathrm{and} \;  Ma = 0.875$. Note that the error bars for the skin friction refer to the uncertainty in the experimental measurements.    }
    
    \label{fig:CfCpDTCSMResults}
\end{figure}

Given that the premise of WMLES involves coarse grids near the wall, which do not resolve the energy-containing turbulent structures, the subgrid-scale stresses are now influenced by the anisotropic mean Reynolds stresses in the first-few cells above the wall. This is exacerbated when there are strong pressure gradients in the flow, and as such, a model that accounts for this anisotropy might be beneficial. In this section, the dynamic tensor coefficient Smagorinsky model (DTCSM) \cite{agrawal2022non} together with the equilibrium wall model is used. In Fig. \ref{fig:CfCpDTCSMResults}, we compare the skin friction and wall pressure coefficient measurements with respect to the experiments. The skin-friction predictions with this model in the upstream region, and in the favorable pressure gradient region are worse than the ones from the DSM. The prediction of the location and the extent of the separation bubble slightly improves with DTCSM over DSM. Although not shown, when a similar suite of numerical simulations was performed for the higher Reynolds number case of the Bachalo-Johnson experiment \cite{bachalo1986transonic} with an extended upstream development of the turbulent boundary layer, and it was observed that the DTCSM results compared more favorably with the experiment than the ones with DSM. This discrepancy between the results of the Sandia ATB and the Bachalo-Johnson case suggests that the upstream history of the incoming boundary layer onto the bump, which had a longer development length in the Bachalo-Johnson case as compared to the Sandia ATB influences the predictions, and that the DTCSM is more sensitive to history effects than the DSM. 

\section{Addressing upstream history effects on the incoming boundary layer}

It was hypothesized that the transition mechanism due to the presence of a small backward-facing step, far upstream of the bump may not be well captured in the simulations due to severe under-resolution. Even at the finest grid considered, the backward-facing step is marginally resolved with approximately three points across its height. This severe under-resolution may lead to incorrect energization of the boundary layer, which may affect the transition location, and perhaps attain the incorrect Reynolds number at the bump. In the authors' own experience, similar challenges had to be overcome for accurate wall-modeled LES of higher Reynolds number Bachalo Johnson experiment. Further, the inviscid  flow acceleration and the development of the boundary layer over the model nose may also be inaccurately captured in coarse wall-modeled LES.  

\begin{figure}[!ht]
    \centering
    \includegraphics[width=0.69\textwidth]{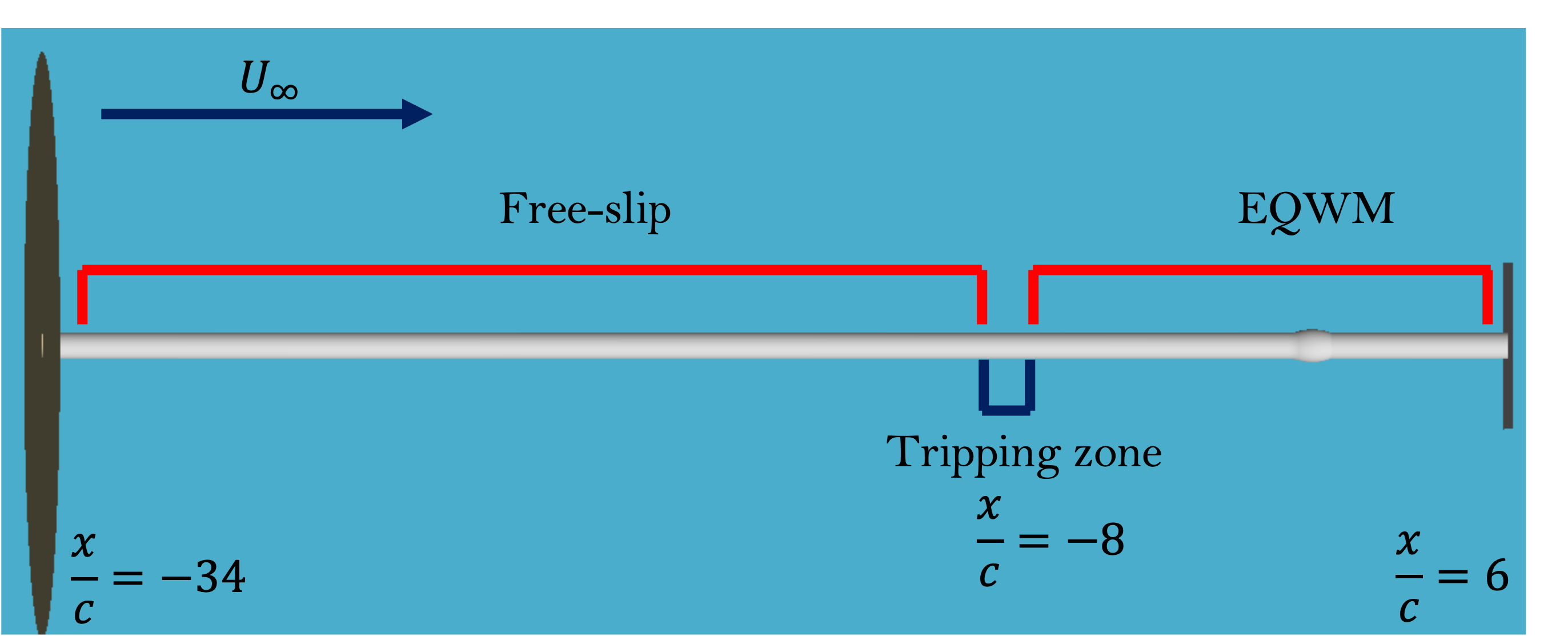}
    \caption{  A schematic of the boundary conditions as applied on the cylinder and the bump model after removing the nose and the backward-facing step. Note that the cylinder is extended upstream compared to the experiment, up to the inlet. The left circular cross-section denotes the inlet, and the right circular cross-section denotes the outlet. The apex of the bump is maintained at $x/c=0.5$. The wind tunnel geometry is replicated from the experiments, and treated with inviscid, free-slip conditions. The tunnel is not shown in this schematic. }
    
    \label{fig:newbcschematic}
\end{figure}

To study the sensitivity of the results to these effects, both the model nose and the backward-facing step were removed, with the cylindrical model extended upstream up to the wind tunnel inlet. The wind tunnel geometry was unchanged. While this may possibly lead to additional inviscid flow acceleration, the area of the cylinder is small (approximately 2\% of the tunnel inlet area) compared to the wind tunnel. To force transition, a suction, and blowing strip is placed upstream of the location of the backward-facing step, to allow the boundary layer to reach the correct skin friction and be free of numerical artifacts by the time it encounters the bump. This strip is forced to sinusoidally vary the wall-normal  velocity with an amplitude comparable to the local friction velocity (estimated from zero-pressure gradient correlations of boundary layers), and a timescale governing the viscous diffusion across the local height of the boundary layer. Free-slip boundary conditions are maintained on the extended cylindrical model until  the point of transition $x/c \sim -8$. A schematic of the application of the boundary conditions in the revised simulation domain is provided in Fig. \ref{fig:newbcschematic}.

\begin{figure}[!ht]
    \centering
    \subfigure[]{
    \includegraphics[width=0.48\textwidth]{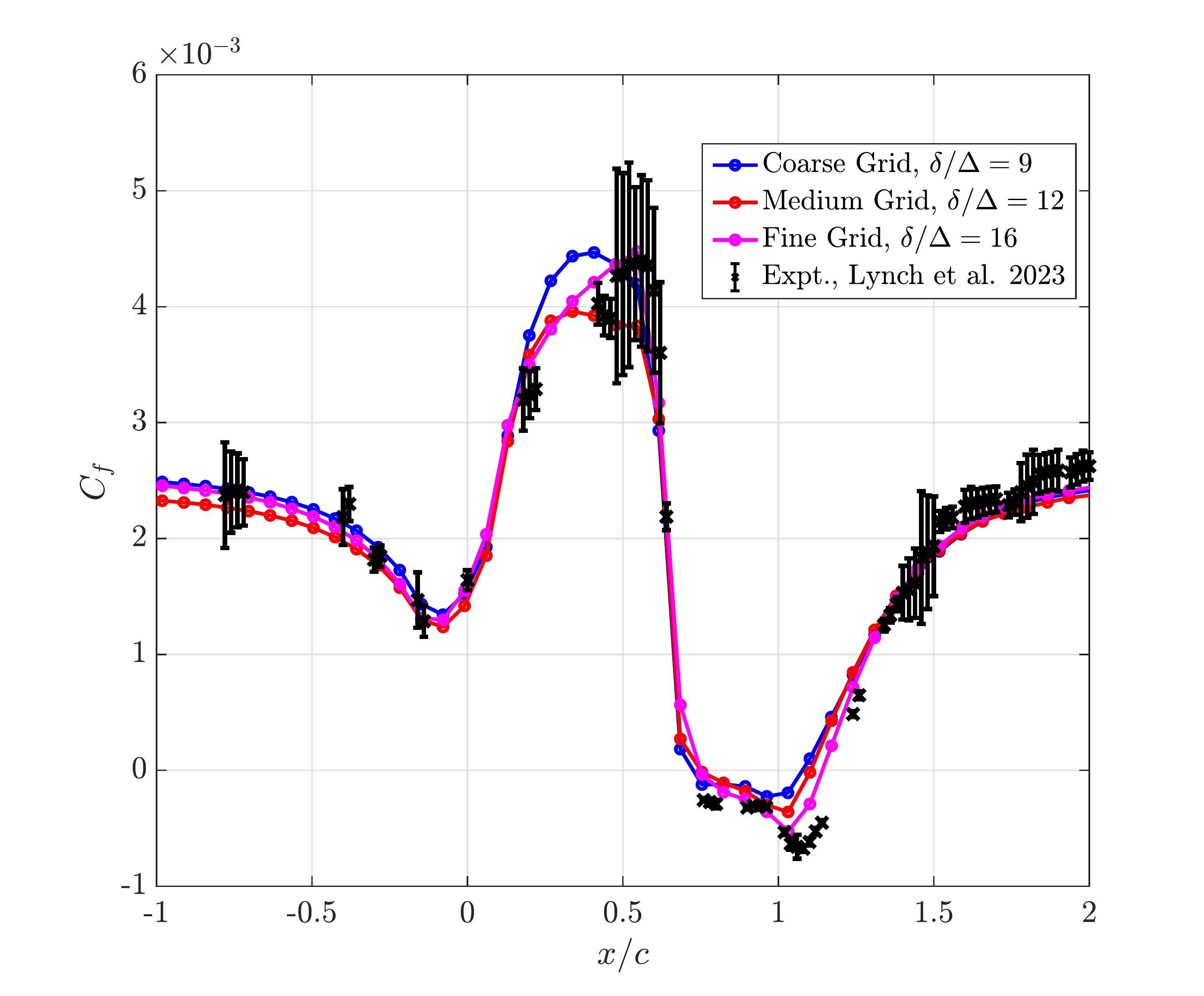}}
    \subfigure[]{
    \includegraphics[width=0.49\textwidth]{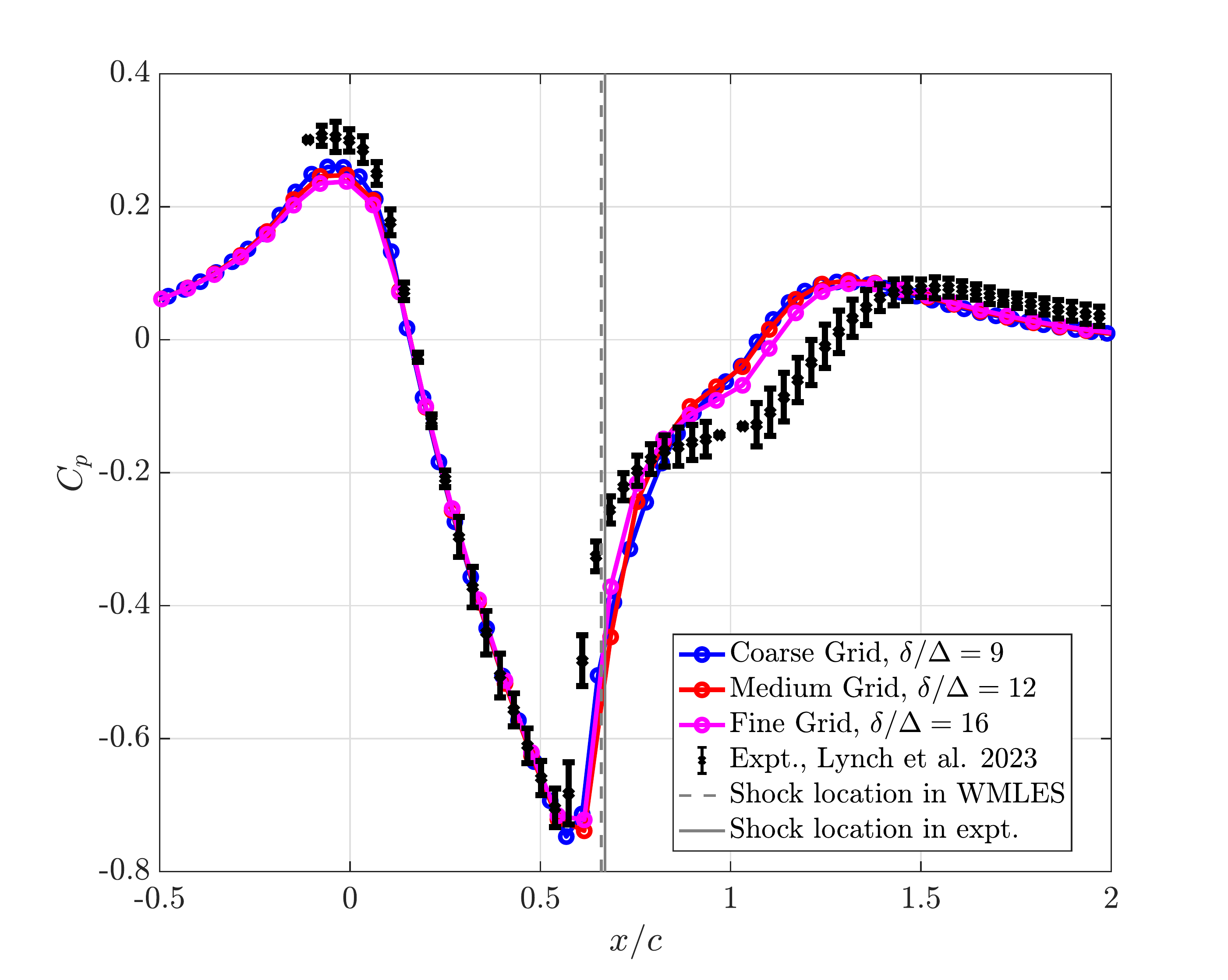}} 
    \caption{  The distribution of (a) Skin friction and (b) wall pressure coefficients for the revised simulation setup, using the dynamic Smagorinsky subgrid-scale and equilibrium wall model. Note that the error bars for the skin friction refer to the uncertainty in the experimental measurements.    }
    
    \label{fig:CfCpDSMnewResults}
\end{figure}

Using the same dynamic modeling procedures of DSM and EQWM, and the same gridding practices as discussed in Section \ref{sec:mesh}, a grid refinement study is performed. Since the cylinder is now extended upstream, the near-wall refinement on the cylinder increases the total number of degrees of freedom to approximately 100 Mil., 300 Mil., and 1.2 Billion, respectively. Fig. \ref{fig:CfCpDSMnewResults} presents the results for $C_f$ and $C_p$ in the revised simulation setup. It is clear that the upstream skin-friction profiles are in much closer agreement with the experiment across all grids. This confirms the hypothesis that the upstream errors  were responsible for the overprediction of skin friction slightly upstream ($-1 \leq x/c \leq 0$) of the bump. 

The behavior of the skin friction in the favorable pressure gradient region and the near the apex is also more favorable with respect to the experiments. The fine grid compares well with the experiment in almost the entirety of the region, $0 < x/c < 0.5$. The resolution on this grid in the region, $0 < x/c < 0.5$ averaged at about $\Delta y_w ^+ \sim 20$, which might indicate that the outer LES starts to capture the flow dynamics well, and that the role of the wall model may be diminishing. Finnicum and Hanratty \cite{finnicum1988effect} showed a significant variation in the mean streamwise velocity profile (above $y^{+} \sim 20$) for a boundary layer under the action of favorable pressure gradient, at $K \sim 2 \times 10^{-6}$. It is then expected that using an equilibrium-type wall model at large $y^{+}$ values, typical in WMLES, would overpredict the wall shear stress, in agreement with the observations in the coarse and medium grids. Although not shown,  on comparing the mean streamwise velocity profiles at $x/c \sim 0.25$ (the location at which the solutions at the three grid-resolutions begin to diverge from one another), it was clear that the three LES solutions reasonably agreed with one another in the outer part of the boundary layer. However, only the fine grid produced a $C_f$ which is in agreement with the experiments, thus confirming that the errors are primarily due to the inaccuracies in the wall model when using coarse resolutions.

Regarding the prediction downstream of the shock, both the $C_f$ and $C_p$ plots suggest that in the region of separation, the grid refinement improves the separation bubble prediction monotonically; with the shock location being predicted correctly in comparison to the experiments. However, even on the finest grid, the separation bubble is underpredicted. Although not shown, using the tensor coefficient subgrid-scale model did not produce qualitatively different results. Further, since the upstream quantities are largely accurate on the fine grid, this deficiency would point to a deficiency of the equilibrium wall model in regions of strong adverse pressure gradients. In strong adverse pressure gradients, for attached boundary layers, the mean streamwise velocity profile shifts upward with a diminished log layer but a stronger wake region. Thus the assumption of a logarithmic velocity profile in outer LES is significantly challenged. In the next section, a recently developed pressure gradient inclusive wall model is applied to this flow to examine these effects.

\section{Improvements in the separation prediction using a pressure-gradient inclusive wall model}
\vspace{4pt}
Agrawal et al. \cite{agrawalarb2022} recently developed a sensor-aided equilibrium wall model which improved the separation prediction on the subsonic flow over the  Boeing speed bump. This model leverages the knowledge of the pressure gradients, and the instantaneous subgrid stresses in the vicinity of, and inside the separation bubble. Simpson \cite{simpson1983model} postulated that in a separated flow, the mean backflow scales with a pressure-gradient-based velocity which competes with the viscous velocity scale. The two scales are given as, 
\begin{equation}
     \vec{u_{\tau, i}} \sim  (\frac{\mu}{\rho} \frac{\partial U_i}{\partial n})^{1/2}   
\hspace{10pt}
\mathrm{and} 
 \hspace{10pt}
\vec{u_{p,i}} \sim (\frac{\mu}{\rho^2} \frac{\partial P}{ \partial x_i})^{1/3},
\end{equation}
where $n$ denotes the wall-normal direction, and $i \; \in  \; \{1,2,3\}$. In the vicinity of a separation point, it is likely that the magnitude of the pressure gradient-based velocity scaling dominates over the skin-friction scaling, and the orientation of the pressure gradient is adverse. Under these conditions, the EQWM may predict excessively low wall stress since the eddy viscosity is tuned for a flow under equilibrium. Thus, for non-equilibrium flow, an additional component of the subgrid shear stress needs to be modeled.

This non-equilibrated, fluctuating stress on the wall is modeled through an instantaneous Neumann extrapolation of the subgrid stresses in the wall-adjacent cell (${\partial \tau^{sgs}_{ij}}/{\partial \hat{n_j}} = 0$, where $\hat{n_j}$ is the $j${th} component of the wall-normal unit vector directed into the fluid). Thus, the wall model can be summarized as 
\begin{equation}
    \tau_{w,i} = 
    \begin{cases}
       \tau^{EQWM}_{w,i}  & \text{if}\ \overline{|\vec{u_p} |} < \overline{|\vec{u_{\tau}}| }, \\
       \tau^{EQWM}_{w,i} - \tau^{sgs}_{ij} \cdot n_j  & \text{otherwise},
    \end{cases}
\end{equation}
where $i,j \;  \in \;  \{1,2,3\}$ and $\tau_{w,i}^{EQWM}$ denotes the wall stress predicted from the EQWM in the $i${th} direction. The operator $\overline{(\cdot )}$ denotes averaging (with an exponential memory) along the homogeneous direction. 

\begin{figure}[!ht]
    \centering
    \subfigure[]{
    \includegraphics[width=0.48\textwidth]{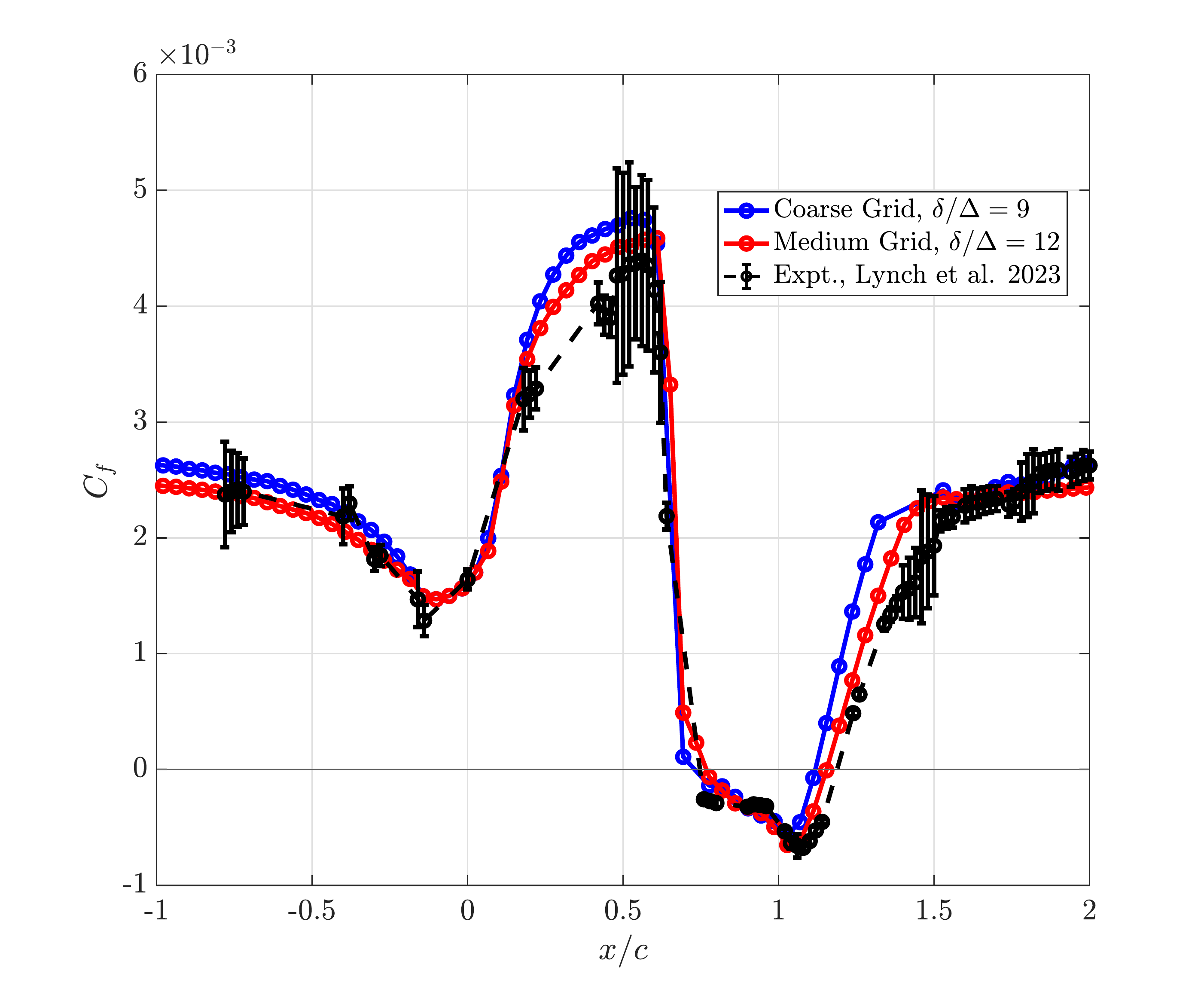}}
    \subfigure[]{
    \includegraphics[width=0.49\textwidth]{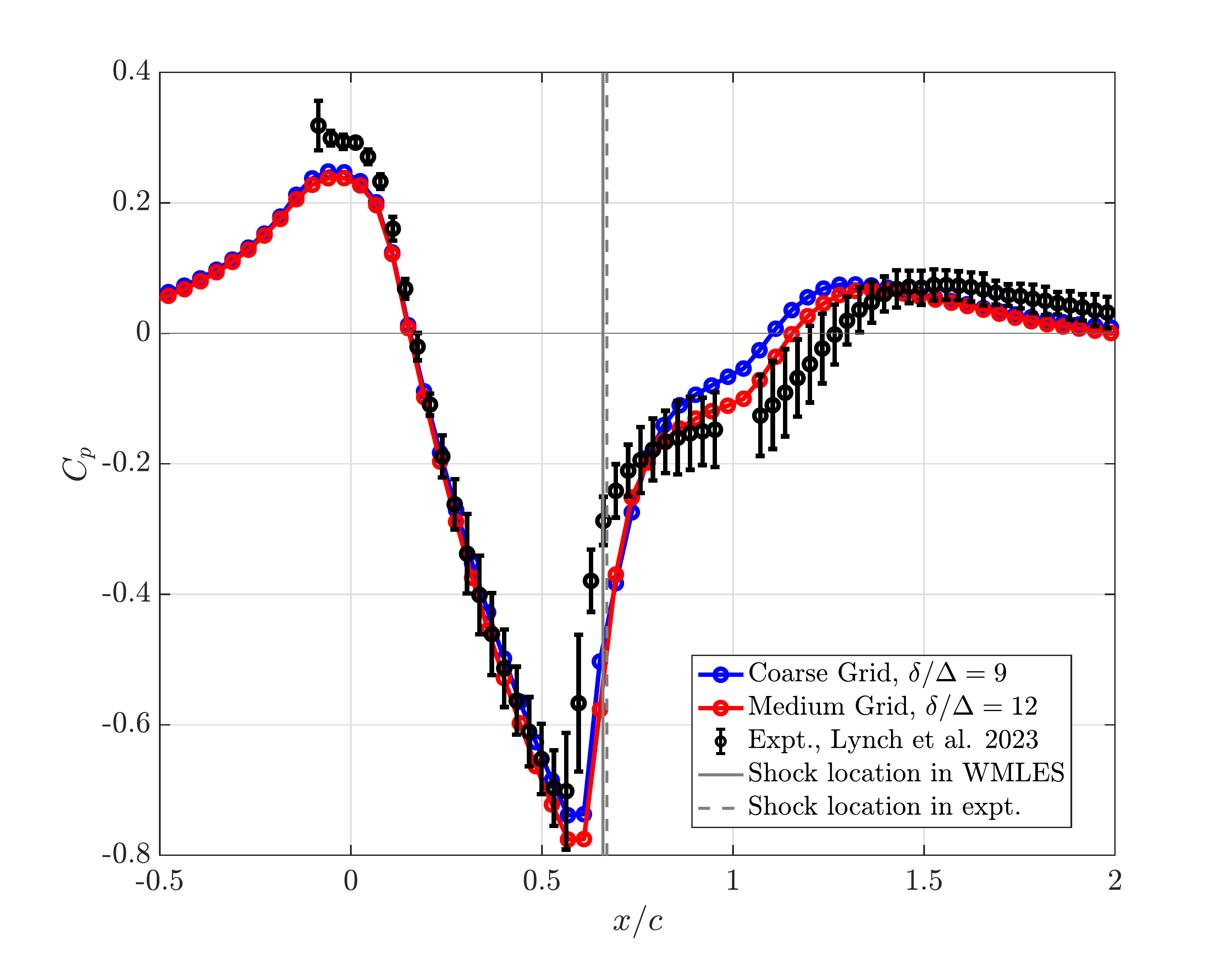}} 
    \caption{  The distribution of (a) Skin friction and (b) wall pressure coefficients with the revised simulation setup, using the dynamic tensor coefficient Smagorinsky subgrid-scale and sensor-aided equilibrium wall model. Note that the error bars for the skin friction refer to the uncertainty in the experimental measurements.    }
    
    \label{fig:CfCpDTCSMSENSORnewResults}
\end{figure}

Agrawal et al. \cite{agrawalarb2022} reported improved predictions from this wall model, when used in conjunction with the tensor-coefficient Smagorsinky subgrid-scale model. The same modeling combination is now used, and the results are reported in Fig. \ref{fig:CfCpDTCSMSENSORnewResults}. The simulations approach the experimental $C_f$ and $C_p$ monotonically on the two grids. It is encouraging that the prediction of $C_f$ on the medium grid (that is 4x cheaper in the total number of control volumes than the fine grid) is comparable to the results obtained on the fine grid with the standard DSM and EQWM models. In fact, the flattening observed in the $C_p$ data is more in agreement with the experiments compared to the fine grid results with DSM and EQWM (see Figure \ref{fig:CfCpDSMnewResults}). Consistent with the DSM and EQWM results, the shock location is in excellent agreement with the experiments.

\section{Conclusion}
In this work, wall-modeled large-eddy simulations are presented for the flow over the Sandia ATB at $Re_c = 1\times 10^6 \; \mathrm{and} \; Ma = 0.875$. On simulating the experimental domain, with both the dynamic Smagorinsky subgrid-scale model and the recently proposed dynamic tensor coefficient Smagorinsky model (DTCSM), the skin friction and wall pressure coefficients are accurately predicted around the region of separation. However, upstream of the bump, the skin friction was overpredicted for both models on all grids considered. This effect was hypothesized to be due to an incorrect prediction of the inviscid acceleration beyond the model nose, and the incorrect energization of the boundary layer due to the under-resolution of the backward-facing step upstream of the bump. Removing the nose and the step, and forcing the flow transition at $x/c \sim -8$ using a synthetic tripping mechanism, provided more favorable comparisons with the experiment in the upstream region, due to the more realistic incoming boundary layer onto the bump. Successive grid refinement allowed for the accurate prediction of $C_f$, and $C_p$ in the favorable pressure gradient of the flow (where the Reynolds shear stresses decrease as the local boundary layer thins, and the flow becomes less resolved). The minimum resolution to achieve this result with the standard modeling approach of the dynamic Smagorinsky model and the equilibrium wall model required approaching the wall to within $20$ viscous units, where the effect of the pressure gradient is directly carried by the resolved flow instead of being accounted for by a wall model. In order to achieve the same accuracy on coarser grids, a pressure-gradient inclusive wall model is  used, and the equilibrium wall stress is augmented by adding the anisotropic subgrid-scale stresses in the regions of strong pressure gradients. These simulations produced results of similar accuracy, as observed on the finest grid with DSM and EQWM, using a 4x (in a total number of control volumes) in both the upstream favorable pressure gradient region as well as within the separation bubble. Overall, even though there are remaining questions on improving the modeling paradigms in the presence of pressure gradients in LES, the results presented in this work are still in much better agreement with the experiments than existing simulation-based studies of this flow. 

\section*{Acknowledgments}
This work was supported by NASA's Transformational Tools and Technologies project under grant number 80NSSC20M0201. Computing resources were awarded through the Oak Ridge Leadership Computing Facility (DoE ALCC) and through the Advanced Simulation and Computing (ASC) program of the US Department of Energy’s National Nuclear Security Administration (NNSA) via the PSAAP-III Center at Stanford, Grant No. DE-NA0002373. The authors acknowledge helpful discussions with Dr. Sanjeeb Bose.  


\end{document}